\documentclass[conference]{IEEEtran}
\IEEEoverridecommandlockouts
\usepackage{cite}
\usepackage{amsmath,amssymb,amsfonts}
\usepackage{algorithmic}
\usepackage{graphicx}
\usepackage{textcomp}
\usepackage{xcolor}

\usepackage{graphicx}
\usepackage{lscape}
\usepackage{lscape}
\usepackage{multirow}
\usepackage{longtable}

\def\BibTeX{{\rm B\kern-.05em{\sc i\kern-.025em b}\kern-.08em
    T\kern-.1667em\lower.7ex\hbox{E}\kern-.125emX}}
\begin{document}

\title{Effective Feature Extraction for Intrusion Detection System using Non-negative Matrix Factorization and Univariate analysis\\
}

\author{\IEEEauthorblockN{1\textsuperscript{st} Swapnil S. Mane}
\IEEEauthorblockA{\textit{Department of Computer Engineering} \\
\textit{College of Engineering}\\
Pune, India \\
maness19.comp@coep.ac.in}
\and
\IEEEauthorblockN{2\textsuperscript{nd} Vaibhav K. Khatavkar}
\IEEEauthorblockA{\textit{Department of Computer Engineering} \\
\textit{College of Engineering}\\
Pune, India \\
vkk.comp@coep.ac.in}
\and
\IEEEauthorblockN{3\textsuperscript{rd} Niranjan R. Gijare}
\IEEEauthorblockA{\textit{Department of Quality and Assurance} \\
\textit{kPoint Technologies Pvt. Ltd.}\\
Pune, India \\
niranjan.gijare@kpoint.com}
\and
\IEEEauthorblockN{4\textsuperscript{th} Pranav V. Bhendawade}
\IEEEauthorblockA{\textit{Software Engineer} \\
\textit{Tata Consultancy Services}\\
Pune, India \\
pranav.bhendawade@tcs.com}
}

\maketitle

\begin{abstract}
An Intrusion detection system (IDS) is essential for avoiding malicious activity. Mostly, IDS will be improved by machine learning approaches, but because of more headers (or features) present in the packet (each record), the model's efficiency is degrading. The proposed model extracts practical features using Non-negative matrix factorization and chi-square analysis.  The more number of features increases the exponential time and risk of overfitting the model. Using both techniques, the proposed model makes a hierarchical approach that will reduce the feature’s quadratic error and noise. The proposed model is implemented on three publicly available datasets, which gives significant improvement. According to recent research, the proposed model has achieved 4.66\% and 0.39\% with respective NSL-KDD and CICD 2017.
\end{abstract}


\section{Introduction}\label{intro}
In the current scenario, there is a tremendous use of Internet networks in that new malicious activity is also created by attackers. So, we resolve this type of problem using an IDS, which will monitor each network to detect ill-disposed activity and detect the new unauthenticated activity. We can observe the primary approach of the IDS in each analysis, which becomes more focused upon machine learning methods to reach many achievable predictions of attack and specific types of attack. The other one is feature extraction that helps the machine learning models improve the possible prediction of attacks. This paper presents an effective and specific feature extraction by machine learning technique, which enhances the IDS utilizing a built-in classifier of machine learning. Also, discover the mentioned challenges of features for IDS, experimental, and implementation evaluation. 
The foremost goal is to obtain a false alarm rate and much good prediction accuracy performance. The first is to build a new model, and the second is to select only appropriate correlated features. So, we build an effective feature extraction method to enhance the efficiency of IDS using machine learning. The article as mentioned above is prepared as follows. Background information present within the next section \ref{back}, section \ref{FE} is a detailed description of features extraction followed by univariate feature extraction in section \ref{UFE} and Non-negative matrix factorization explained in section \ref{NFE}. Followed by, recommended feature extraction technique introduced in section \ref{model}, afterward feature extraction using Non-negative matrix factorization is in section and building blocks of the proposed model explained in same section. proposed experiment details in section\par
The outcomes of the recommended technique and comparison of recent research done in this field are inside section \ref{exp}. Ultimately, section \ref{conl} presents a specific conclusion of the proposed method. 

\section{Background information}\label{back}
\subsection{Feature Extraction}\label{FE}
Feature extraction is a dimensionality reduction method. At a specific initial state, noisy data is decomposed to more achievable feature vectors for further processing.   Feature extraction is an essential technique in data mining that will catch relevant features from a dataset. Dataset is in fixed numbers of features that will be in continuous or categorical, binary format. Feature extraction is nothing but recognizing valid information is much specific to the domain and will be more likely to the possible measurements. The features may forward packet details that will be informed in the network traffic part, such as, e.g., protocol, flag, service, flow duration, etc. Manually, we try to transform raw features into valuable features that will be more complicated than automatic feature extraction methods. There will be irrelevant features in our data, then obviously decrease the efficiency of many machine learning models. Also, the advantages of feature extraction are to reduce overfitting that will be irrelevant data or low opportunity data to make decisions based on noisy data, minimize training and testing time and improve accuracy.

\subsection{Univariate feature extraction}\label{UFE}
Univariate selected feature strongly related to responsible variables based on the univariate test that will assign the score to individual features, based on the score they select top features \cite{2}. In sci-kit learn library gives the 'SelectKBest' class that will use with a bunch of various statistical analyses to choose an appropriate number of features. In our proposed method we use that chi-squared statistical test to extract the features. This technique is simple to execute and best for achieving a good understanding of data.

\begin{table}[htbp]
\centering
\caption{Scores of features by the univariate feature extraction method}
\label{univariat}
\begin{tabular}{|l|l|l|}
\hline
 & \textbf{Features} & \textbf{Score} \\ \hline
0 & duration & 212.600148 \\ \hline
3 & flag & 148.645727 \\ \hline
6 & wrong\_fragment & 68.071177 \\ \hline
17 & num\_outbound\_cmds & 40.089442 \\ \hline
1 & protocol\_type & 39.644819 \\ \hline
9 & logged\_in & 38.838046 \\ \hline
16 & num\_access\_files & 22.672851 \\ \hline
7 & hot & 16.333728 \\ \hline
2 & service & 15.882329 \\ \hline
21 & srv\_count & 13.800332 \\ \hline
5 & dst\_bytes & 13.724378 \\ \hline
\end{tabular}
\end{table}

\subsection{Non-negative feature extraction}
\label{NFE}
Non-negative Matrix Factorization (NMF) makes a matrix analysis approach that will describe every matrix in the below format

\begin{equation}
X_{pxq} = W_{pxr} * H_{rxq} \label{eq}
\end{equation}
NMF is decomposed a given data (X) in two matrices (H and W) that hold the original information in a unique product of pair matrices. The detailed methodology is elaborated in followed proposed feature extraction method.

\section{Proposed model}\label{model}
This section describes the recommended feature extraction method, which includes two foremost techniques: feature extraction using NMF and then Univariate Feature extraction. We assume that our dataset is in a non-negative numeric format. Some dataset has Nan and Infinity cells that will be replaced by or resolved by simple imputer this a method of preprocessing in machine learning. Features X is passed through the recommended method besides different variables, and they are elaborated here.\\
\begin{itemize}

\item A specific number of components for Non- negative matrix factorization will help as an output number of features, U.
\item A particular number of components for univariate feature extraction serves as unique features to the model, V.
\item A specific number of features is obtained by univariate, which will generate a unique.\\
\end{itemize}
Feature space that will be major suitable during classification.

NMF is to be implemented in various machine learning difficulties while clustering and feature extraction. In this recommended method, a distinct strategy under this Non-negative matrix factorization will combine features to make unique feature space, and they will be more diminutive than the original feature space. Afterward, we assign W and H using Non-negative matrix factorization proposed by Gallopoulos E \cite{3}. In our proposed method, we divide the primary features into groups then merge every group feature in a single space. This number of collected groups will be equal to the user-defined components (U) chosen for NMF, and this is the first parameter.\par
\begin{figure}[htbp]
\centerline{\includegraphics[width=80mm,height=35mm]{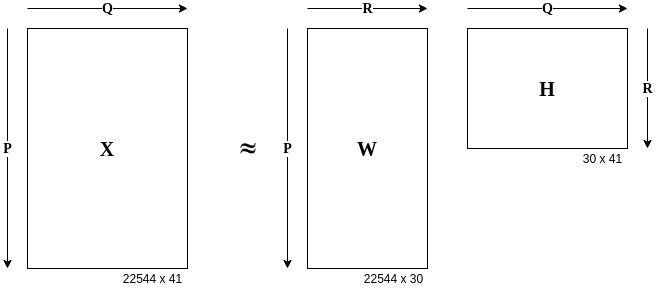}}
\caption{Non-negative matrix factorization for feature extraction}
\label{nmf}
\end{figure}

The new features by NMF consist of U feature vectors merged in the n sized vector, which will new feature space. The example of a Kaggle dataset that is a military environment dataset for the intrusion detection system being factorized into 30 components (U = 30) is shown in fig ~\ref{nmf}.\\

\begin{figure}[htbp]
\centerline{\includegraphics[width=85mm,height=75mm]{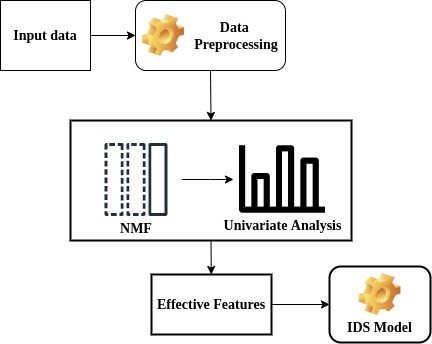}}
\caption{Flow of proposed model}
\label{propose}
\end{figure}

\begin{figure}[htbp]
\centerline{\includegraphics[width=40mm,height=30mm]{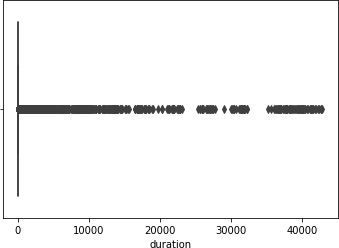}}
% \end{figure}
% \begin{figure}[htbp]
\centerline{\includegraphics[width=40mm,height=30mm]{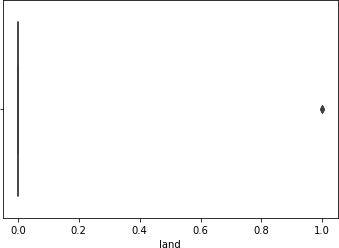}}
\caption{Distribution of quantitative data of land and duration feature}
\label{dist}
\end{figure}

Later the feature extraction by applying Non- negative matrix factorization, the recommended method attempts to obtain highly correlated features. The proposed method feature extraction using univariate, which will normalize the features. Also, a summary of the recommended method is shown in fig. \ref{propose}. In this flow, as usual, we load the complete dataset that includes noisy data. The features and labels separately store two distinguished variables, i.e., X and y, as shown in the flow chart. Then describe all features, which will give count, mean, min, max, and other factors for each feature, manually Drop that features there mean it will be approx. Zero. The reason for dropping the feature is that the fig. \ref{dist} duration feature has a high average that will be highly correlated. On the other hand, feature name land has nearby zero average shown in fig. \ref{dist}, and the land feature is approx. Zero means that is not much impact on the classification model. This is a manual feature extraction method. Dataset has ‘Nan’ and ‘Infinity.’\par

cells that are present that will resolve by the simple imputer method of Scikit-Learn \cite{5}, cells replaced by the average of that particular feature. Afterward, label encoding of y in that name of the attack is replaced by a numeric value that will help the machine learning algorithm better train and predict the model. As features have numeric values, but that will be very high and small values that will impact the model so that will convert into Term Frequency-Inverse Document Frequency (TF-IDF) is a statistical evaluation, how relevant a word (single value) is to a document (a record) in a collection of documents (records). This will be done by multiplication of two-term: how many times a value appears in a record and the inverse document frequency of the value across a set of records. Then feature reduction by NMF in the kind that will help to the univariate feature extraction method, that will take filtered features by NMF. After, univariate feature extraction gets high score features that the chi-squared statistical hypothesis will test.

\section{Experiment Details}\label{exp}
In our proposed method, we use several datasets to compare some recent research done in this field. Comparison is made with the help of the accuracy of the metrics of several models of machine learning. The dataset used in our proposed experiment is provided in Table \ref{dataset}.

\begin{table}[htbp]
\centering
\caption{Description of the dataset used for the implementation}
\label{dataset}
\begin{tabular}{|l|l|l|l|}
\hline
\textbf{Dataset} & \textbf{Records} & \textbf{classes} & \textbf{Reference} \\ \hline
\begin{tabular}[c]{@{}l@{}}Military Network \\ Environment from Kaggle\end{tabular} & 25,192 & 2 & {[}10{]} \\ \hline
\begin{tabular}[c]{@{}l@{}}UNB Intrusion Detection \\ Evaluation CICID 2017 (Web Attack)\end{tabular} & 170,366 & 4 & {[}11{]} \\ \hline
20\% training data of NSL–KDD & 25,192 & 22 & {[}12{]} \\ \hline
\end{tabular}
\end{table}

The dataset of the military network environment from Kaggle is to be audited, which consists of an intrusion in a network of a military environment. It acquired dump TCP/IP packets for a military network by simulating a United states Airforce LAN. This dataset simulates the number of attacks in a binary format that is an anomaly and normal. For every TCP/IP connection, 41 Qualitative and quantitative features were obtained from the anomaly and normal data. The classes in data are two categories and their respective records Anomalous (11743) and Normal (13449). CICID 2017 famous dataset for an intrusion detection system. We use only the Web attack morning working hours Thursday dataset. This web attack dataset included four categories Benign (168186), Brute force attack (1507), XSS attack (652), SQL Injection (21). NSL-KDD dataset is the improvement of the KDD’99 dataset, and this is also a popular dataset for an IDS. It has twenty-two categories usual (67343),  ipsweep (3599), smurf (2646), teardrop (892), pod (201), buffer\_overflow (30), land (18), rootkit (10), multihop (7), phf (4),spy (2), warezmaster (20), ftp\_write	(8), Neptune	(41214),	satan (3633),warezclient (890), guess\_passwd (53), port sweep (2931), Nmap (1493), IMAP (11), back (956), loadmodule (9), Perl (3) Our experiment was done in jupyter notebook using python we used the following libraries: Scikit-Learn [5], pandas, Numpy [4] and seaborn

\begin{figure}[htbp]
\centerline{\includegraphics[width=65mm,height=70mm]{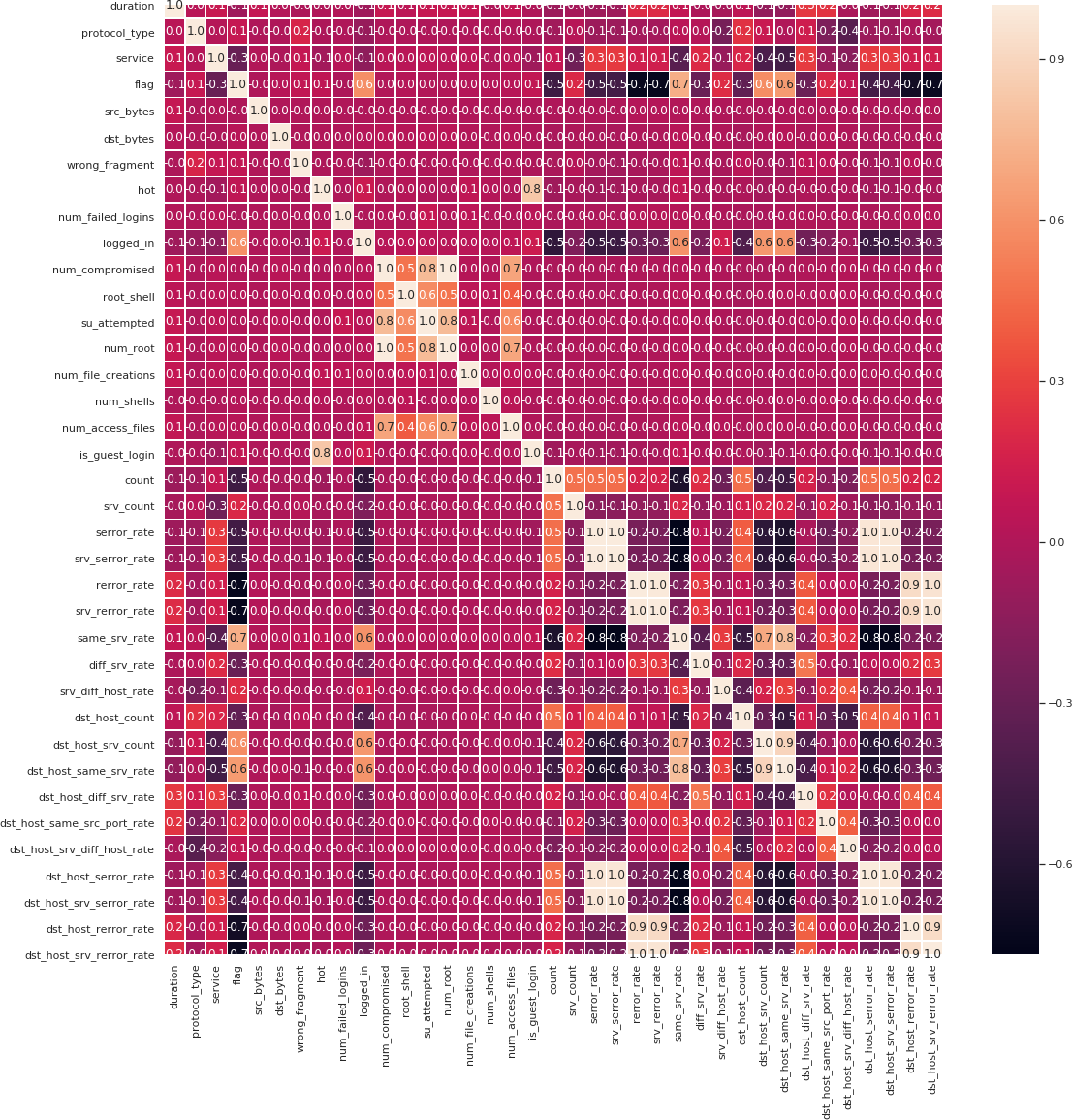}}
\caption{Correlation matrix before feature extraction}
\label{corr1}
\end{figure}

To observe all correlations between the military network environment features, we visualize this matrix by heatmap of the seaborn method. The color value of the cells is proportionate to the number of measurements that match specific dimensional values.\\

\begin{figure}[htbp]
\centerline{\includegraphics[width=65mm,height=70mm]{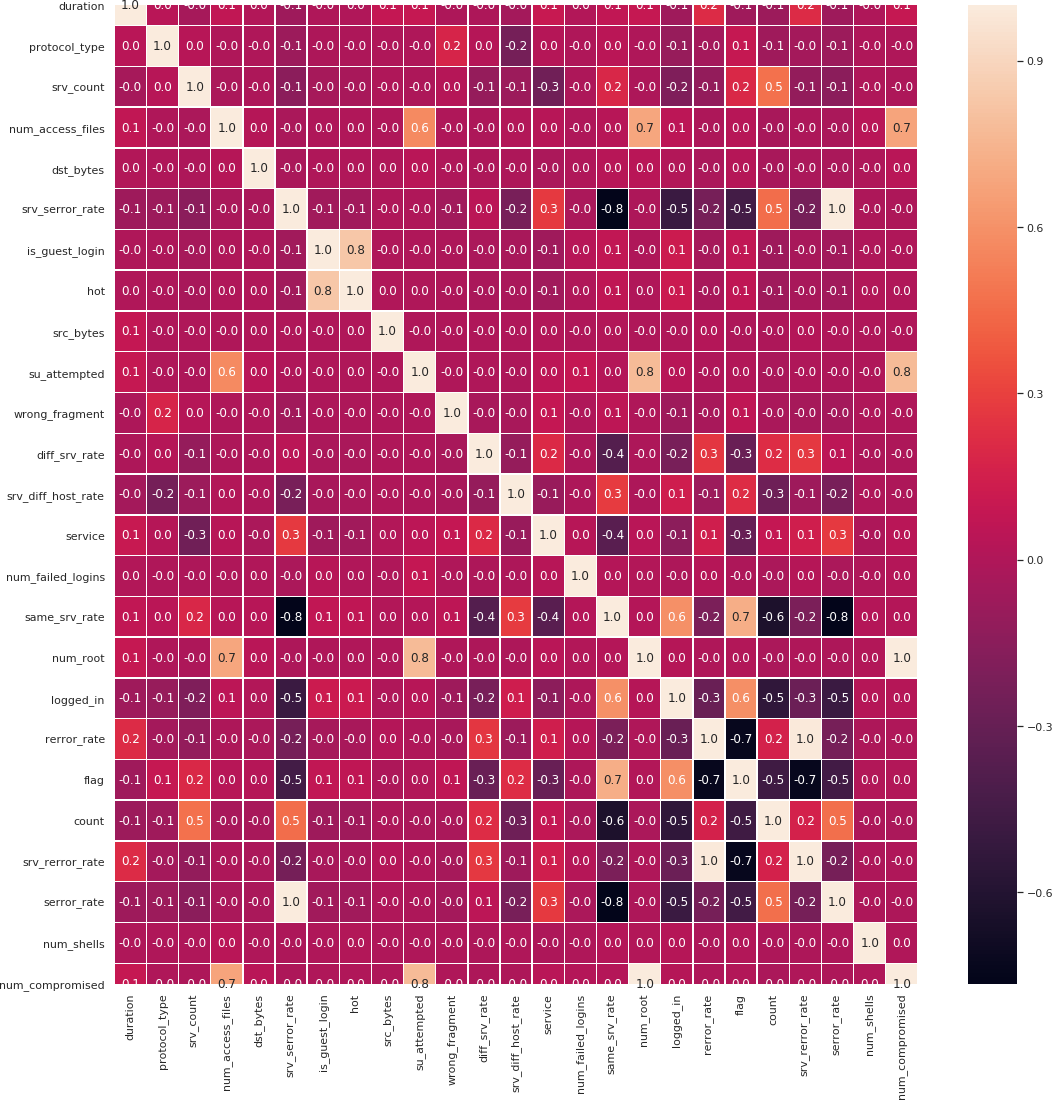}}
\caption{Correlation matrix after a proposed feature extraction method}
\label{corr2}
\end{figure}

In observation, after the Feature extraction obtained the highly correlated features after the proposed feature extraction is shown in \ref{corr1} and \ref{corr2} before and after feature extraction. After comparing them, our proposed feature extraction method to achieve highly efficient features. Here also presents a significant result of this recommended method. Results achieved by the proposed approach will compare to the recent research done in this field
% Please add the following required packages to your document preamble:
% \usepackage{multirow}
% \usepackage{graphicx}

\begin{figure}[htbp]
\centerline{\includegraphics[width=90mm,height=70mm]{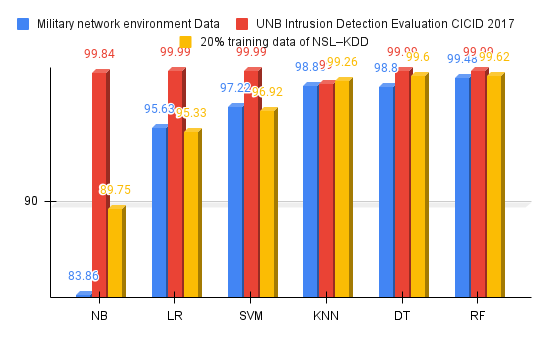}}
\caption{Results of the proposed model}
\label{chart1}
\end{figure}

\begin{figure}[htbp]
\centerline{\includegraphics[width=90mm,height=70mm]{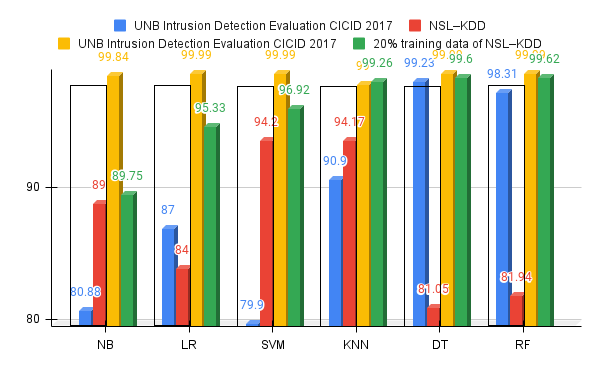}}
\caption{Correlation matrix after a proposed feature extraction method}
\label{chart2}
\end{figure}

The accuracy of various machine learning algorithms improved after the proposed feature extraction. There is no research done on the dataset of the Military Network Environment from Kaggle. The paper has presented all recently enhanced machine learning algorithms with a respective dataset of NSL-KDD and CICID 2017. Fig. \ref{chart1} shows the performance of the traditional machine learning algorith using the proposed method. The Naive Bayes, Linear Regression, Support Vector Machine, K Nearest Neighbor, Decision Tree, and Random Forest models are abbreviated as NB, LR, SVM, KNN, DT, and RF. In fig. \ref{chart2} shows a comparison between the proposed model with [6], [8], [9] and [12]. The black-bordered bars are recent research performances, and the remaining are proposed performances.\\

\section{Conclusion}\label{conl}
Information security is a critical problem for an organization and all fields related to the internet, and it leads to stealing money and personal information. This will resolve by the intrusion detection system, so while we stated, Feature extraction performs a much essential component in various programmed systems, i.e., intrusion detection system, classification, decision making, and the like. Also, advanced research is done in this area, specifically effective feature extraction but is not limited to combining two specific feature extraction methods.\\
The proposed paper presented an effective feature extraction for a classification model that will be collect highly correlated features from a more significant number of noisy features. We use the exiting technique as univariate and NMF, build a proposed model. This method combines two feature extraction methods, and the indusial method extracts feature differently to obtain highly correlated filtered features. After the proposed effective feature extraction, it will reduce raw data and time of execution, the accuracy of any model that will be little or much improved than feature extraction done by any single technique. According to recent research, the proposed model has achieved 4.66\% and 0.39\% with respective NSL-KDD and CICD 2017.

\end{document}